# Commuting and Noncommuting Coordinates in Deformed Special Relativity

Clemens Heuson

*Ulrichstr. 1, D-87493 Lauben, Germany*
e-mail: *clemens.heuson@freenet.de*

Commuting and noncommuting space-time coordinates in a class of deformed special relativity theories are investigated. Their momentum space representation, transformation behaviour, space-time algebra, invariants and the corresponding field theories are derived. Several coordinates require as a novel feature the introduction of deformed plane waves.

## 1. Introduction

Deformed or doubly special relativity (DSR) with two invariant scales, a velocity scale *c* and a fundamental length scale $l$ or energy scale $\kappa = 1/l$ presumably related to the Planck length or mass, has been discussed widely in the last years as a possible extension of special relativity (SR) to the high energy region, see the reviews in [1]. A main problem in DSR theories certainly is the proper formulation in space-time, see [1] and recent discussions in [2]. In [3] it was shown by using a Hamiltonian formalism that noncommuting coordinates lead to a simple transformation law and invariant metric together with a proper definition of velocity, while commuting coordinates obey a more complicated transformation law, see also [4,5]. However by using noncommuting coordinates one is forced to employ noncommutative field theory, which is more difficult especially in the context of DSR [6].

Concerning the introduction of space time coordinates in DSR theories [1-12] one immediately faces two questions: the first one is their commutativity, the second one is their transformation behaviour, while both are related to the nonlinear transformation of momenta. This gives several options for the coordinates: commuting or noncommuting with deformed or undeformed transformation behaviour. We investigate their properties in a class of DSR theories with energy independent speed of light and discuss their relation to some previous suggestions in the literature. In the context of DSR theories one has to navigate somehow between the Scylla of noncommuting coordinates with simple transformations and more difficult field theory and the Charybdis of commuting coordinates with complicated transformations but hopefully simpler commutative field theory. Of course one would like to investigate the impact of DSR theories on quantum field theory.

## 2. DSR in Momentum Space

A possible starting point of any DSR theory is a modified dispersion relation written in the form

$$F^2 E^2 - G^2 \mathbf{p}^2 = m^2 \tag{1a}$$

with the functions $F, G(E, \mathbf{p}^2, l)$ preserving rotational symmetry. From this one derives the deformed Lorenz transformations in momentum space leaving invariant the relation (1). Depending on the functions *F*, *G* a multitude of dispersion relations is possible. In this letter we focus on the simple case *G=F*, where the equations can be written in a rather compact form. Equation (1a) then becomes

$$F^2 p^2 + m^2 = 0 \tag{1b}$$

We use $\hbar = c = 1$, the metric $\text{diag}(\eta) = (-,+,+,+)$ and else the notations in [5]. Examples are the Magueijo-Smolin model [7] and the model in [8], see also [4],[9]. The transformation of momenta leaving invariant (1b) is from [5]

$$p'^{\mu} = A\Lambda^{\mu}_{\nu} p^{\nu}, \quad p'_{\mu} = A\overline{\Lambda}^{\nu}_{\mu} p_{\nu}, \quad p^{\mu} = A'\overline{\Lambda}^{\mu}_{\nu} p'^{\nu}, \quad p_{\mu} = A'\Lambda^{\nu}_{\mu} p'_{\nu}, \quad F' = F/A \tag{2}$$

where $\Lambda^\mu_\nu$ is the standard Lorentz transformation and $\bar{\Lambda}^\nu_\mu$ it's inverse with velocity $v \to -v$ obeying $\Lambda^\mu_\rho \bar{\Lambda}^\rho_\nu = \delta^\mu_\nu$ and $\eta_{\mu\nu}\Lambda^\mu_\rho\Lambda^\nu_\sigma = \eta_{\rho\sigma}$. The last equation in (2) is derived from the invariance of the modified dispersion relation. The function $A = A(p^0, p^1, v)$ can be calculated in the various models and $A' = A(p'^0, p'^1, -v)$ with $A' = 1/A$ from (2). The transformation of momentum derivatives following from (2) was derived in [5] as

$$\frac{\partial}{\partial p'^\mu} = \frac{\partial p^\nu}{\partial p'^\mu}\frac{\partial}{\partial p^\nu} = \frac{1}{A}\bar{\Lambda}^\nu_\mu(\frac{\partial}{\partial p^\nu} - a_\nu p^\lambda \frac{\partial}{\partial p^\lambda}) , \quad a_\mu = \frac{A_\mu}{A + p^\lambda A_\lambda} , \quad A_\mu = \frac{\partial A}{\partial p^\mu} \qquad (3)$$

We now write down several quantities for the models [7],[8] used here and in following sections in the compact form of the following table.

Table 1. Quantities in specific DSR models

| Model [7] | Model [8] |
|---|---|
| $F = (1 - l\, p^0)^{-1}$ | $F = (1 - l^2 \boldsymbol{p}^2)^{-1/2}$ |
| $A = (1 - l\, p^0 + l\gamma(p^0 - v\, p^1))^{-1}$ | $A = (1 - l^2 p^{0\,2} + l^2\gamma^2(p^0 - v\, p^1)^2)^{-1/2}$ |
| $a_\mu = (l(1-\gamma), l\gamma v, 0, 0)$ | $a_\mu = (l^2\gamma^2 v(p^0 - v\, p^1), l^2\gamma^2 v(p^1 - v\, p^0), 0, 0)$ |
| $b_\mu = l\delta_{\mu 0}$ | $b_\mu = l^2 p_k \delta_{\mu k}$ |
| $B_{0i} = -l p_i, B_{ij} = 0$ | $B_{0i} = -l^2 p^0 p_i, B_{ij} = 0$ |
| $1 - p^\lambda b_\lambda = 1/F$ | $1 - p^\lambda b_\lambda = 1/F^2$ |
| $1 - p^\lambda a_\lambda = 1/A$ | $1 - p^\lambda a_\lambda = 1/A^2$ |

## 3. Noncommuting DSR Coordinates

In [10] coordinates were introduced demanding the invariance of the contraction $\pi \cdot \xi = p \cdot x$, where $\pi$ and $\xi$ are the auxiliary SR variables representing momenta and coordinates. From $\pi = F \cdot p$ one gets immediately $\xi = (1/F) \cdot x$. Rewriting the auxiliary commuting coordinates $\xi$ in physical momenta gives [5]

$$\xi_\mu = i\frac{\partial}{\partial \pi^\mu} = \frac{i}{F}(\frac{\partial}{\partial p^\mu} - b_\mu p^\lambda \frac{\partial}{\partial p^\lambda}) , \quad b_\mu = \frac{F_\mu}{F + p^\lambda F_\lambda} , \quad F_\mu = \frac{\partial F}{\partial p^\mu} \qquad (4)$$

They transform under standard Lorentz formations $\xi'_\mu = \bar{\Lambda}^\nu_\mu \xi_\nu$. The expressions for $b_\mu$ in the models [7],[8] can be found in table 1. Thereby the above introduced coordinates $\hat{x}_\mu$ (which we here denote with a hat in difference to [5]) in momentum space are:

$$\hat{x}_\mu = F \cdot \xi_\mu = i(\frac{\partial}{\partial p^\mu} - b_\mu p^\lambda \frac{\partial}{\partial p^\lambda}) \qquad (5)$$

These coordinates however are noncommuting contrary to the implicit assumption in [10] but in agreement with [3],[4],[5], as may be derived from their momentum representation in (5). Their deformed Lorentz transformations from $F' = F/A$ and $\xi'_\mu = \bar{\Lambda}^\nu_\mu \xi_\nu$ are surprisingly simple

$$\hat{x}'^\mu = \frac{1}{A}\Lambda^\mu_\nu \hat{x}^\nu , \quad \hat{x}'_\mu = \frac{1}{A}\bar{\Lambda}^\nu_\mu \hat{x}_\nu , \quad \hat{x}^\mu = \frac{1}{A'}\bar{\Lambda}^\mu_\nu \hat{x}'^\nu , \quad \hat{x}_\mu = \frac{1}{A'}\Lambda^\nu_\mu \hat{x}'_\nu \qquad (6)$$

The transformation of noncommuting coordinate derivatives is obtained from (6) with $A' = 1/A$

$\frac{\partial}{\partial \hat{x}'^\mu} = \frac{\partial \hat{x}^\nu}{\partial \hat{x}'^\mu}\frac{\partial}{\partial \hat{x}^\nu} = A\bar{\Lambda}^\nu_\mu \frac{\partial}{\partial \hat{x}^\nu} , \quad \frac{\partial}{\partial \hat{x}'_\mu} = \frac{\partial \hat{x}_\nu}{\partial \hat{x}'_\mu}\frac{\partial}{\partial \hat{x}_\nu} = A\Lambda^\mu_\nu \frac{\partial}{\partial \hat{x}_\nu}$. Thereby the momentum operator, which must (!) transform like (2), can be expressed through noncommuting coordinate derivatives as

$$p^\mu = -i\frac{\partial}{\partial \hat{x}_\mu} = -i\hat{\partial}^\mu \tag{7}$$

With (7) the dispersion relation (1b) can be expressed in the models [7],[8] in terms of these derivatives and gives after multiplying with $\phi$ the deformed Klein Gordon equations

$$\left(\frac{-\hat{\partial}^\mu \hat{\partial}_\mu}{(1+il\hat{\partial}^0)^2} + m^2\right)\phi(\hat{x}) = 0 \; [7] \;,\; \left(\frac{-\hat{\partial}^\mu \hat{\partial}_\mu}{1+l^2\hat{\partial}_k^2} + m^2\right)\phi(\hat{x}) = 0 \; [8] \tag{8}$$

The algebra between coordinates and momenta derived from (5) is

$$[\hat{x}_\mu, p_\nu] = i(\eta_{\mu\nu} - b_\mu p_\nu) \,,\; [\hat{x}_\mu, \hat{x}_\nu] = i(b_\mu \hat{x}_\nu - b_\nu \hat{x}_\mu) \,,\; [p_\mu, p_\nu] = 0 \tag{9}$$

The invariants under the transformations (2),(6) are

$$F^2 p^\mu p_\mu = inv,\, p^\mu \hat{x}_\mu = inv,\, s^2 = (1/F^2)\hat{x}^\mu \hat{x}_\mu = inv \tag{10}$$

The invariant metric is in accordance with the rainbow metric in [10], where however commuting coordinates were assumed. The deformed boost and rotation generators $M_{\mu\nu}$ obeying the standard algebra are obtained in momentum space as

$$M_{\mu\nu} = p_\nu \hat{x}_\mu - p_\mu \hat{x}_\nu = i(p_\nu \frac{\partial}{\partial p^\mu} - p_\mu \frac{\partial}{\partial p^\nu} + B_{\mu\nu}\, p^\lambda \frac{\partial}{\partial p^\lambda}) \tag{11}$$

where $B_{\mu\nu} = p_\mu b_\nu - p_\nu b_\mu$. Due to the assumed rotational symmetry $F = F(p^0, \vec{p}^2, l)$ only the boost generators $M_{0i}$ are deformed. Assuming that the coordinates are commuting in a first approximation, one obtains field theories with higher order derivatives considered in [11].

Another interesting possibility mentioned in [5] is to consider noncommuting coordinates $\hat{y}_\mu = (1/F^2)\hat{x}_\mu$ transforming like momenta as the coordinates do in SR. Their momentum space representation is

$$\hat{y}_\mu = \frac{1}{F^2}\hat{x}_\mu = \frac{i}{F^2}(\frac{\partial}{\partial p^\mu} - b_\mu p^\lambda \frac{\partial}{\partial p^\lambda}) \tag{12}$$

They transform like momenta in DSR $\hat{y}'_\mu = A\overline{\Lambda}_\mu^\nu \hat{y}_\nu$. The space-time algebra is then

$$[\hat{y}_\mu, p_\nu] = \frac{i}{F^2}(\eta_{\mu\nu} - b_\mu p_\nu) \,,\; [\hat{y}_\mu, \hat{y}_\nu] = -\frac{i}{F^2}(b_\mu \hat{y}_\nu - b_\nu \hat{y}_\mu) \,,\; [p_\mu, p_\nu] = 0) \tag{13}$$

The invariants requiring deformed plane waves are

$$F^2 p^\mu p_\mu = inv,\, F^2 p^\mu \hat{x}_\mu = inv,\, s^2 = F^2 \hat{x}^\mu \hat{x}_\mu = inv \tag{14}$$

From $\frac{\partial}{\partial \hat{y}_\mu} = \frac{\partial \hat{x}_\nu}{\partial \hat{y}_\mu}\frac{\partial}{\partial \hat{x}_\nu} = F^2 \frac{\partial}{\partial \hat{x}_\mu}$ or short $\breve{\partial}^\mu = F^2 \hat{\partial}^\mu$ and $p^\mu = -i\hat{\partial}^\mu$ one derives for the model in [7] $\breve{\partial}^\mu = \hat{\partial}^\mu / (1+il\hat{\partial}^0)^2$ and for the model in [8] $\breve{\partial}^\mu = \hat{\partial}^\mu / (1+l^2(\hat{\partial}^k)^2)$. Solving for $\hat{\partial}^\mu$ gives finally

$$p^\mu = -i\hat{\partial}^\mu = -i\left(\frac{1-\sqrt{1-4il\,\breve{\partial}^0}}{2il\,\breve{\partial}^0}\right)^2 \breve{\partial}^\mu \; [7],\; p^\mu = -i\hat{\partial}^\mu = -i\frac{1-\sqrt{1-4il^2(\breve{\partial}^k)^2}}{2l^2(\breve{\partial}^k)^2}\breve{\partial}^\mu \; [8] \tag{15}$$

The deformed Klein Gordon equations are therefore

$$\left(-\left(\frac{1-\sqrt{1-4i\,l\hat{\partial}^0}}{2i\,l\hat{\partial}^0}\right)^2 \hat{\partial}^\mu \hat{\partial}_\mu + m^2\right)\Phi = 0 \ [7], \ \left(-\frac{1-\sqrt{1-4i\,l^2(\hat{\partial}^k)^2}}{2l^2(\hat{\partial}^k)^2}\hat{\partial}^\mu \hat{\partial}_\mu + m^2\right)\Phi = 0 \ [8] \tag{16}$$

If one defines the deformed generators as

$$M_{\mu\nu} = p_\nu \hat{x}_\mu - p_\mu \hat{x}_\nu = F^2(p_\nu \hat{y}_\mu - p_\mu \hat{y}_\nu) \tag{17}$$

then their algebra remains standard. Using $M_{\mu\nu} = p_\nu \hat{y}_\mu - p_\mu \hat{y}_\nu$ would result in a deformed Lorentz algebra. These coordinates also yield a higher order noncommutative field theory.

## 4. Commuting DSR Coordinates

In this section we investigate commuting coordinates with deformed transformation behaviour. The most simple suggestion is to take the derivatives with respect to momenta as commuting coordinates, which we here denote by $x_\mu$ without hat:

$$x_\mu = i\frac{\partial}{\partial p^\mu} \tag{18}$$

The connection of the commuting coordinates $x_\mu$ with the noncommuting coordinates $\hat{x}_\mu$ and its inversion, obtained by contracting with $p^\mu$, is from (5) and (18)

$$\hat{x}_\mu = x_\mu - b_\mu p^\lambda x_\lambda, \ x_\mu = \hat{x}_\mu + \frac{F_\mu}{F} p^\lambda \hat{x}_\lambda \tag{19}$$

The boost and rotation generators become with (11) $M_{\mu\nu} = p_\nu x_\mu - p_\mu x_\nu + B_{\mu\nu} p^\lambda x_\lambda$, and the algebra between coordinates and momenta is of course canonical. We note the behaviour of the commuting coordinates $x_\mu$ under deformed Lorentz transformations, which can be obtained from their momentum space representation in (18) and (3)

$$x'_\mu = \frac{1}{A}\bar{\Delta}_\mu^\nu x_\nu = \frac{1}{A}\bar{\Lambda}_\mu^\rho(\delta_\rho^\nu - a_\rho p^\nu)x_\nu, \ x'^\mu = \frac{1}{A}\Delta_\nu^\mu x^\nu = \frac{1}{A}\Lambda_\rho^\mu(\delta_\nu^\rho - a^\rho p_\nu)x^\nu \tag{20}$$

Since $A = A(p^0, p^1)$ only $a_0, a_1 \neq 0$. One realizes that the transformation law is more complicated than for noncommuting coordinates, however in both cases a dependence on momenta is unavoidable. The matrix $\bar{\Delta}$ is explicitly given by

$$\bar{\Delta} = \begin{pmatrix} \gamma(1 - p^0(a_0 + va_1)) & \gamma(v - p^1(a_0 + va_1)) & -\gamma p^2(a_0 + va_1) & -\gamma p^3(a_0 + va_1) \\ \gamma(v - p^0(a_1 + va_0)) & \gamma(1 - p^1(a_1 + va_0)) & -\gamma p^2(a_1 + va_0) & -\gamma p^3(a_1 + va_0) \\ 0 & 0 & 1 & 0 \\ 0 & 0 & 0 & 1 \end{pmatrix} \tag{21}$$

It's determinant is $det(\bar{\Delta}) = 1 - p^\rho a_\rho = A/(A + p^\lambda A_\lambda)$ which can be evaluated in the models [7],[8], with results shown in table 1. Thereby one obtains the Jacobian $J$ of the commuting coordinates $d^4 x' = J d^4 x$ as $J = det(\partial x'_\mu/\partial x_\nu) = det(\bar{\Delta})/A^4$.

We evaluate the transformation matrix $\bar{\Delta}$ with the general expression (21) in model [7]

$$\overline{\Delta} = \begin{pmatrix} \gamma - \mathrm{l}\, p^0(\gamma-1) & \gamma v - \mathrm{l}\, p^1(\gamma-1) & -\mathrm{l}\, p^2(\gamma-1) & -\mathrm{l}\, p^3(\gamma-1) \\ \gamma v - \mathrm{l}\, p^0 \gamma v & \gamma - \mathrm{l}\, p^1 \gamma v & -\mathrm{l}\, p^2 \gamma v & -\mathrm{l}\, p^3 \gamma v \\ 0 & 0 & 1 & 0 \\ 0 & 0 & 0 & 1 \end{pmatrix} \qquad (22)$$

It is in full agreement with the matrix derived in [3] from a Hamiltonian formalism for commuting coordinates in 1+1 dimensions, noticing $x_\mu = (-t, x_i)$ and $v \to -v$ from the different convention for momentum Lorentz transformations. The commuting coordinates considered in [3] therefore are given by (18) in momentum space. For the determinant one obtains $det(\overline{\Delta}) = 1/A$ which gives $J = 1/A^5$. In model [8] we get for the same matrix

$$\overline{\Delta} = \begin{pmatrix} \gamma - \mathrm{l}^2 p^0 p^1 \gamma v & \gamma v - \mathrm{l}^2 p^1 p^1 \gamma v & -\mathrm{l}^2 p^1 p^2 \gamma v & -\mathrm{l}^2 p^1 p^3 \gamma v \\ \gamma v - \mathrm{l}^2 p^0 p^0 \gamma v & \gamma - \mathrm{l}^2 p^0 p^1 \gamma v & -\mathrm{l}^2 p^0 p^2 \gamma v & -\mathrm{l}^2 p^0 p^3 \gamma v \\ 0 & 0 & 1 & 0 \\ 0 & 0 & 0 & 1 \end{pmatrix} \qquad (23)$$

The determinant becomes $det(\overline{\Delta}) = 1/A^2$ and $J = 1/A^6$. Interestingly this matrix agrees with the DSR Lorentz matrix in [9] with the same conventions as above. Thereby one concludes that the canonical coordinates introduced in [9], where the question of their commutativity was not discussed, are identical to the commuting coordinates $x_\mu$ in (18). The invariant volume element in coordinate space is obtained from the Jacobian $J$ together with $F' = F/A$ in model [7] as $d^4x/F^5$ and in model [8] as $d^4x/F^6$.

Finally we look at the invariants built from the commuting coordinates $x_\mu$. First observe that the contraction $p \cdot x$ is not invariant, instead one gets from (2) and (20) $p'^\mu x'_\mu = (1 - p^\rho a_\rho) p^\mu x_\mu$. The invariant combination is obtained from $p^\mu \hat{x}_\mu$ as

$$(1 - p^\lambda b_\lambda) p^\mu x_\mu = inv \qquad (24)$$

yielding deformed plane waves, see table 1 for the corresponding expressions in [7],[8]. The invariant metric $s^2 = 1/F^2 \, \hat{x}^\mu \hat{x}_\mu$ can be rewritten in commuting coordinates with (19) giving a more complicated invariant homogeneous quadratic form, the momentum invariant is the same as in (10). From (19) and $\frac{\partial}{\partial x_\mu} = \frac{\partial \hat{x}_\nu}{\partial x_\mu} \frac{\partial}{\partial \hat{x}_\nu} = \frac{\partial}{\partial \hat{x}_\mu} - p^\mu b_\nu \frac{\partial}{\partial \hat{x}_\nu}$ with $p^\mu = -i \hat{\partial}^\mu$ one gets $\partial^\mu = \hat{\partial}^\mu (1 + i b_\nu \hat{\partial}^\nu)$. In model [7] this gives $\partial^\mu = \hat{\partial}^\mu (1 + i\, \mathrm{l}\, \hat{\partial}^0)$ and in model [8] $\partial^\mu = \hat{\partial}^\mu (1 + \mathrm{l}^2 (\hat{\partial}^k)^2)$, which can be solved for $\hat{\partial}^\mu$. Thereby one gets expressions for the momentum operator in terms of commuting coordinate derivatives resulting in higher order field equations. In model [7] and similarly in model [8] (in a first order approximation, the exact value can be derived with Cardano's formula) one obtains finally for the momentum operator

$$p^\mu = -i \hat{\partial}^\mu = \frac{-i \partial^\mu}{\tfrac{1}{2}(1 + \sqrt{1 + 4i\, \mathrm{l}\, \partial^0})} \; [7], \quad p^\mu = -i \hat{\partial}^\mu \approx \frac{-i \partial^\mu}{1 + \mathrm{l}^2 (\partial^k)^2} \; [8] \qquad (25)$$

Inserting in (1b) gives after multiplying with the scalar field $\phi$ in both cases rather complicated higher order deformed Klein Gordon equations expressed by derivatives with respect to the commuting DSR coordinates.
The main problems with these commuting DSR coordinates are the complicated expressions for the transformation behaviour and the coordinate representation of the momentum operator.

## 5. Commuting and Noncommuting SR Coordinates

Another option for commuting coordinates in DSR theories was already mentioned in [5] and investigated in [12], but the necessity, to introduce deformed plane waves was not mentioned. One can take the auxiliary SR variables from (4) $\xi_\mu = (1/F)\hat{x}_\mu$ as commuting coordinates.

$$\xi_\mu = i\frac{\partial}{\partial \pi^\mu} = \frac{i}{F}(\frac{\partial}{\partial p^\mu} - b_\mu p^\lambda \frac{\partial}{\partial p^\lambda}), \ \xi'_\mu = \bar{\Lambda}_\mu^\nu \xi_\nu \tag{26}$$

The algebra between coordinates and momenta is then given by

$$[\xi_\mu, p_\nu] = \frac{i}{F}(\eta_{\mu\nu} - b_\mu p_\nu), \ [\xi_\mu, \xi_\nu] = 0, \ [p_\mu, p_\nu] = 0 \tag{27}$$

Thus the coordinates $\xi_\mu$ transform according the standard Lorentz transformations in (26), while the momenta $p^\mu$ transform under the deformed Lorentz transformations in (2). The boost and rotation generators by using $\hat{x}_\mu = F\xi_\mu$ in (11) are $M_{\mu\nu} = F(p_\nu \xi_\mu - p_\mu \xi_\nu)$ and as for (11) the commutators between them remain standard. The invariants built from these quantities are

$$F^2 p^\mu p_\mu = inv, \ F\, p^\mu \xi_\mu = inv, \ \xi^\mu \xi_\mu = inv \tag{28}$$

Similar to the last section we get deformed plane waves with these commuting SR coordinates

$$\phi = N \cdot \exp(i F p^\mu \xi_\mu) \tag{29}$$

Now we want to derive a representation of the momentum operator in coordinate space, which only can be done in an explicit model. Consider first model [7]. From $\frac{\partial}{\partial \xi^\mu} = \frac{\partial \hat{x}^\nu}{\partial \xi^\mu}\frac{\partial}{\partial \hat{x}^\nu} = F\frac{\partial}{\partial \hat{x}^\mu}$ or shortly $\slashed{\partial}_\mu = F\hat{\partial}_\mu$ and with $F = 1/(1+l\, p_0)$ and $p_\mu = -i\hat{\partial}_\mu$ one obtains $\slashed{\partial}_\mu = \hat{\partial}_\mu/(1-i\, l\, \hat{\partial}_0)$. Solving for $\hat{\partial}_\mu$ gives finally for the momentum operator in terms of commuting coordinate derivatives in model [7] and similarly in model [8]

$$p_\mu = \frac{-i\slashed{\partial}_\mu}{1+i\, l\, \slashed{\partial}_0} \ [7], \quad p_\mu = \frac{-i\slashed{\partial}_\mu}{\sqrt{1-l^2 \slashed{\partial}_k^2}} \ [8] \tag{30}$$

Inserting them in the deformed dispersion relation $F^2 p^2 + m^2 = 0$ applied to a scalar field, one obtains in both cases the standard wave equation $(-\slashed{\partial}^\mu \slashed{\partial}_\mu + m^2)\phi(\xi) = 0$, which should not come out as great surprise, since this is exactly the expression invariant under standard Lorentz transformations. Inserting the deformed plane wave (29) in this equation gives again the modified dispersion relation (1b). The theory is nevertheless different from standard field theory, since the momenta transform under (2) and thereby plane waves must be modified according (29). Modified plane waves have been considered in a Lorentz invariant setting based on a generalized uncertainty principle within a first order approximation in [13]. The problems concerning coordinates in DSR discussed recently in [2] do not exist for the commuting SR coordinates $\xi_\mu$, since there is no momentum dependence in their transformations.

A generalized uncertainty principle in the form $[\hat{\xi}_\mu, \pi_\nu] = i\Theta_{\mu\nu}(\pi)$ can lead to noncommuting coordinates. An important example is given by the coordinates $\hat{\xi}_\mu$ transforming according SR [14]

$$\hat{\xi}_\mu = i(\frac{\partial}{\partial \pi^\mu} \pm l^2 \pi_\mu \pi^\lambda \frac{\partial}{\partial \pi^\lambda}) \tag{31}$$

For possible modifications see [15]. The algebra derived from (31) is

$$[\hat{\xi}_\mu,\hat{\xi}_\nu]=\pm i\,l^2(\hat{\xi}_\mu\pi_\nu-\hat{\xi}_\nu\pi_\mu),\ [\hat{\xi}_\mu,\pi_\nu]=i(\eta_{\mu\nu}\pm l^2\pi_\mu\pi_\nu),\ [\pi_\mu,\pi_\nu]=0 \qquad (32)$$

(31) and (32) define the famous Snyder and anti Snyder model. The entire theory is Lorentzinvariant, nevertheless it represents an interesting deformation of special relativity. One could of course rewrite the relations in terms of the physical momenta. However due to the noncommutativity of these coordinates the quantum field theory is as difficult as for the noncommuting DSR coordinates in section 2.

## 6. Coordinates in generalized DSR Models

Recently in [16] a class of DSR models with $F=(1-l^n|\mathbf{p}|^n)^{-1/n}$ was proposed generalizing the model in [8]. Similarly a class with $F=(1-l^n E^n)^{-1/n}$ generalizing the model in [7] may be considered and both classes show the group property. As mentioned in [8], this model may be rewritten in a form, so that it belongs to the second class too for $n=2$. In table 2 we display several expressions used in sections 2-5 in these generalized classes, where $p=(p_x^2+p_y^2+p_z^2)^{1/2}$ and $E=p^0$. The expressions in model [7] are obtained for $n=1$ from the first column in table 2 and in model [8] for $n=2$ from the second column. The various coordinates and momentum operators of the previous sections together with their algebra can be easily obtained for these two classes by substituting the corresponding expressions in table 2.

The invariant volume element $d\mu_x$ for commuting DSR coordinates $x_\mu$ is obtained from $d^4x'=J\,d^4x$ and together with $J=\det(\partial x'_\mu/\partial x_\nu)=(1-p^\rho a_\rho)/A^4=A^{-n-4}$ gives $d\mu_x=F^{-n-4}d^4x$. For the noncommuting DSR coordinates $\hat{x}_\mu$ one gets $J=\det(\partial \hat{x}'_\mu/\partial \hat{x}_\nu)=A^{-4}$ and thereby $d\mu_{\hat{x}}=F^{-4}d^4\hat{x}$. The commuting SR coordinates $\xi_\mu$ of course have an undeformed invariant volume element. The invariant volume element in momentum space is obtained from $d^4p'=J\,d^4p$ with $J=\det(\partial p'^\mu/\partial p^\nu)=A^3(A+p^\rho A_\rho)=A^{n+4}$ and therefore is given by the expression $d\mu_p=F^{n+4}d^4p$.

Table 2. Quantities in generalized DSR Models

| Models generalizing [7] | Models generalizing [8] |
|---|---|
| $F=(1-l^n E^n)^{-1/n}$ | $F=(1-l^n p^n)^{-1/n}$ |
| $A=(1-l^n E^n+l^n\gamma^n(E-v\,p_x)^n)^{-1/n}$ | $A=(1-l^n p^n+l^n(\gamma^2(p_x-vE)^2+p_y^2+p_z^2)^{n/2})^{-1/n}$ |
| $b_\mu=l^n E^{n-1}\delta_{\mu 0}$ | $b_\mu=l^n p^{n-2}p_i\delta_{\mu i}$ |
| $B_{0i}=-l^n E^{n-1}p_i,\ B_{ij}=0$ | $B_{0i}=-l^n p^{n-2}E\,p_i,\ B_{ij}=0$ |
| $1-p^\lambda b_\lambda=F^{-n}$ | $1-p^\lambda b_\lambda=F^{-n}$ |
| $1-p^\lambda a_\lambda=A^{-n}$ | $1-p^\lambda a_\lambda=A^{-n}$ |
| $d\mu_{\hat{x}}=F^{-4}d^4\hat{x}$ | $d\mu_{\hat{x}}=F^{-4}d^4\hat{x}$ |
| $d\mu_x=F^{-n-4}d^4x$ | $d\mu_x=F^{-n-4}d^4x$ |
| $d\mu_\xi=d^4\xi$ | $d\mu_\xi=d^4\xi$ |
| $d\mu_p=F^{n+4}d^4p$ | $d\mu_p=F^{n+4}d^4p$ |

## 7. Summary and Outlook

In summary we derived for several commuting and noncommuting coordinates in a class of DSR theories their momentum space representation, transformation behaviour, space-time algebra, their invariants and the corresponding field theories. Noncommuting DSR coordinates obey a simple transformation law, but yield a noncommutative field theory, which is more difficult. The same is valid for noncommuting coordinates transforming according SR. Commuting DSR coordinates obey a complicated transformation law, require deformed plane waves and the corresponding field theory also is not very simple. Alternatively it is possible to use commuting SR coordinates without momentum dependence in their transformations and a standard field theory, but with deformed plane waves. The relation of the various coordinates to previous suggestions in the literature was analyzed.

First steps to a quantum field theory with commuting DSR coordinates were undertaken in [11], but it was unclear how gauge invariant interactions could be introduced. A possible solution in the context of a generalized uncertainty principle was discussed in [13], which however leads to a rather complicated interaction structure even for abelian gauge fields. For both commuting and noncommuting DSR coordinates one encounters field theories with higher order derivatives. By employing the commuting SR coordinates one can use standard field theory and interactions in coordinate space, but with the complication of deformed plane waves. Weighing the options for the various coordinates by inspecting the corresponding Klein Gordon equations or momentum operators, it seems that this last possibility is the simplest way.